# Generalized roof duality and bisubmodular functions

Vladimir Kolmogorov
University College London
v.kolmogorov@cs.ucl.ac.uk

**Abstract**

Consider a convex relaxation $\hat{f}$ of a pseudo-boolean function $f$. We say that the relaxation is *totally half-integral* if $\hat{f}(\boldsymbol{x})$ is a polyhedral function with half-integral extreme points $\boldsymbol{x}$, and this property is preserved after adding an arbitrary combination of constraints of the form $x_i = x_j$, $x_i = 1 - x_j$, and $x_i = \gamma$ where $\gamma \in \{0, 1, \frac{1}{2}\}$ is a constant. A well-known example is the *roof duality* relaxation for quadratic pseudo-boolean functions $f$. We argue that total half-integrality is a natural requirement for generalizations of roof duality to arbitrary pseudo-boolean functions.

Our contributions are as follows. First, we provide a complete characterization of totally half-integral relaxations $\hat{f}$ by establishing a one-to-one correspondence with *bisubmodular functions*. Second, we give a new characterization of bisubmodular functions. Finally, we show some relationships between general totally half-integral relaxations and relaxations based on the roof duality.

## 1 Introduction

Let $V$ be a set of $|V| = n$ nodes and $\mathcal{B} \subset \mathcal{K}^{1/2} \subset \mathcal{K}$ be the following sets:

$$\mathcal{B} = \{0,1\}^V \qquad \mathcal{K}^{1/2} = \{0, \tfrac{1}{2}, 1\}^V \qquad \mathcal{K} = [0,1]^V$$

A function $f : \mathcal{B} \to \mathbb{R}$ is called *pseudo-boolean*. In this paper we consider convex relaxations $\hat{f} : \mathcal{K} \to \mathbb{R}$ of $f$ which we call *totally half-integral*:

**Definition 1.** *(a) Function $\hat{f} : \mathcal{P} \to \mathbb{R}$ where $\mathcal{P} \subseteq \mathcal{K}$ is called* half-integral *if it is a convex polyhedral function such that all extreme points of the epigraph $\{(\boldsymbol{x}, z) \mid \boldsymbol{x} \in \mathcal{P}, z \geq \hat{f}(\boldsymbol{x})\}$ have the form $(\boldsymbol{x}, \hat{f}(\boldsymbol{x}))$ where $\boldsymbol{x} \in \mathcal{K}^{1/2}$. (b) Function $\hat{f} : \mathcal{K} \to \mathbb{R}$ is called* totally half-integral *if restrictions $\hat{f} : \mathcal{P} \to \mathbb{R}$ are half-integral for all subsets $\mathcal{P} \subseteq \mathcal{K}$ obtained from $\mathcal{K}$ by adding an arbitrary combination of constraints of the form $x_i = x_j$, $x_i = \overline{x}_j$, and $x_i = \gamma$ for points $\boldsymbol{x} \in \mathcal{K}$. Here $i, j$ denote nodes in $V$, $\gamma$ denotes a constant in $\{0, 1, \frac{1}{2}\}$, and $\overline{z} \equiv 1 - z$.*

A well-known example of a totally half-integral relaxation is the *roof duality relaxation* for quadratic pseudo-boolean functions $f(\boldsymbol{x}) = \sum_i c_i x_i + \sum_{(i,j)} c_{ij} x_i x_j$ studied by Hammer, Hansen and Simeone [13]. It is known to possess the *persistency* property: for any half-integral minimizer $\hat{\boldsymbol{x}} \in \arg\min \hat{f}(\hat{\boldsymbol{x}})$ there exists minimizer $\boldsymbol{x} \in \arg\min f(\boldsymbol{x})$ such that $x_i = \hat{x}_i$ for all nodes $i$ with integral component $\hat{x}_i$. This property is quite important in practice as it allows to reduce the size of the minimization problem when $\hat{\boldsymbol{x}} \neq \boldsymbol{\frac{1}{2}}$. The set of nodes with guaranteed optimal solution can sometimes be increased further using the `PROBE` technique [6], which also relies on persistency.

The goal of this paper is to generalize the roof duality approach to arbitrary pseudo-boolean functions. The total half-integrality is a very natural requirement of such generalizations, as discussed later in this section. As we prove, total half-integrality implies persistency.

We provide a complete characterization of totally half-integral relaxations. Namely, we prove in section 2 that if $\hat{f} : \mathcal{K} \to \mathbb{R}$ is totally half-integral then its restriction to $\mathcal{K}^{1/2}$ is a *bisubmodular function*, and conversely any bisubmodular function can be extended to a totally half-integral relaxation.

**Definition 2.** *Function* $f : \mathcal{K}^{1/2} \to \mathbb{R}$ *is called* bisubmodular *if*

$$f(\boldsymbol{x} \sqcap \boldsymbol{y}) + f(\boldsymbol{x} \sqcup \boldsymbol{y}) \quad \leq \quad f(\boldsymbol{x}) + f(\boldsymbol{y}) \qquad\qquad \forall\, \boldsymbol{x}, \boldsymbol{y} \in \mathcal{K}^{1/2} \tag{1}$$

*where binary operators* $\sqcap, \sqcup : \mathcal{K}^{1/2} \times \mathcal{K}^{1/2} \to \mathcal{K}^{1/2}$ *are defined component-wise as follows:*

| $\sqcap$ | $0$ | $\frac{1}{2}$ | $1$ |
|---|---|---|---|
| $0$ | $0$ | $\frac{1}{2}$ | $\frac{1}{2}$ |
| $\frac{1}{2}$ | $\frac{1}{2}$ | $\frac{1}{2}$ | $\frac{1}{2}$ |
| $1$ | $\frac{1}{2}$ | $\frac{1}{2}$ | $1$ |

| $\sqcup$ | $0$ | $\frac{1}{2}$ | $1$ |
|---|---|---|---|
| $0$ | $0$ | $0$ | $\frac{1}{2}$ |
| $\frac{1}{2}$ | $0$ | $\frac{1}{2}$ | $1$ |
| $1$ | $\frac{1}{2}$ | $1$ | $1$ |

(2)

As our second contribution, we give a new characterization of bisubmodular functions (section 3). Using this characterization, we then prove several results showing links with the roof duality relaxation (section 4).

## 1.1 Applications

This work has been motivated by computer vision applications. A fundamental task in vision is to infer pixel properties from observed data. These properties can be the type of object to which the pixel belongs, distance to the camera, pixel intensity before being corrupted by noise, etc. The popular *MAP-MRF approach* casts the inference task as an *energy minimization* problem with the objective function of the form $f(\boldsymbol{x}) = \sum_C f_C(\boldsymbol{x})$ where $C \subset V$ are subsets of neighboring pixels of small cardinality ($|C| = 1, 2, 3, \ldots$) and terms $f_C(\boldsymbol{x})$ depend only on labels of pixels in $C$.

For some vision applications the roof duality approach [13] has shown a good performance [29, 31, 22, 23, 32, 1, 16, 17].[1] Functions with higher-order terms are steadily gaining popularity in computer vision [30, 32, 1, 16, 17]; it is generally accepted that they correspond to better image models. Therefore, studying generalizations of roof duality to arbitrary pseudo-boolean functions is an important task. In such generalizations the **total** half-integrality property is essential. Indeed, in practice, the relaxation $\hat{f}$ is obtained as the sum of relaxations $\hat{f}_C$ constructed for each term independently. Some of these terms can be $c|x_i - x_j|$ and $c|x_i + x_j - 1|$. If $c$ is sufficiently large, then applying the roof duality relaxation to these terms would yield constraints $x_i = x_j$ and $x = \overline{x}_j$ present in the definition of total half-integrality. Constraints $x_i = \gamma \in \{0, 1, \frac{1}{2}\}$ can also be simulated via the roof duality, e.g. $x_i = x_j, x_i = \overline{x}_j$ for the same pair of nodes $i, j$ implies $x_i = x_j = \frac{1}{2}$.

## 1.2 Related work

**Half-integrality** There is a vast literature on using half-integral relaxations for various combinatorial optimization problems. In many cases these relaxations lead to 2-approximation algorithms. Below we list a few representative papers.

The earliest work recognizing half-integrality of polytopes with certain pairwise constraints was perhaps by Balinksi [3], while the persistency property goes back to Nemhauser and Trotter [27] who considered the vertex cover problem. Hammer, Hansen and Simeone [13] established that these properties hold for the *roof duality relaxation* for quadratic pseudo-boolean functions. Their work was generalized to arbitrary pseudo-boolean functions by Lu and Williams [24]. (The relaxation in [24] relied on converting function $f$ to a multinomial representation; see section 4 for more details.) Hochbaum [14, 15] gave a class of integer problems with half-integral relaxations. Very recently, Iwata and Nagano [18] formulated a half-integral relaxation for the problem of minimizing submodular function $f(\boldsymbol{x})$ under constraints of the form $x_i + x_j \geq 1$.

[1] In many vision problems variables $x_i$ are not binary. However, such problems are often reduced to a sequence of *binary* minimization problems using iterative move-making algorithms, e.g. using *expansion moves* [9] or *fusion moves* [22, 23, 32, 17].

In computer vision, several researchers considered the following scheme: given a function $f(\boldsymbol{x}) = \sum f_C(\boldsymbol{x})$, convert terms $f_C(\boldsymbol{x})$ to *quadratic* pseudo-boolean functions by introducing auxiliary binary variables, and then apply the roof duality relaxation to the latter. Woodford et al. [32] used this technique for the stereo reconstruction problem, while Ali et al. [1] and Ishikawa [16] explored different conversions to quadratic functions.

To the best of our knowledge, all examples of totally half-integral relaxations proposed so far belong to the class of *submodular relaxations*, which is defined in section 4. They form a subclass of more general bisubmodular relaxations.

**Bisubmodularity** Bisubmodular functions were introduced by Chandrasekaran and Kabadi as rank functions of *(poly-)pseudomatroids* [10, 19]. Independently, Bouchet [7] introduced the concept of $\Delta$-matroids which is equivalent to pseudomatroids. Bisubmodular functions and their generalizations have also been considered by Qi [28], Nakamura [26], Bouchet and Cunningham [8] and Fujishige [11]. The notion of the *Lovász extension* of a bisubmodular function introduced by Qi [28] will be of particular importance for our work (see next section).

It has been shown that some submodular minimization algorithms can be generalized to bisubmodular functions. Qi [28] showed the applicability of the ellipsoid method. A weakly polynomial combinatorial algorithm for minimizing bisubmodular functions was given by Fujishige and Iwata [12], and a strongly polynomial version was given by McCormick and Fujishige [25].

Recently, we introduced *strongly* and *weakly tree-submodular* functions [21] that generalize bisubmodular functions.

## 2 Total half-integrality and bisubmodularity

The first result of this paper is following theorem.

**Theorem 3.** *If $\hat{f} : \mathcal{K} \to \mathbb{R}$ is a totally half-integral relaxation then its restriction to $\mathcal{K}^{1/2}$ is bisubmodular. Conversely, if function $f : \mathcal{K}^{1/2} \to \mathbb{R}$ is bisubmodular then it has a unique totally half-integral extension $\hat{f} : \mathcal{K} \to \mathbb{R}$.*

This section is devoted to the proof of theorem 3. Denote $\mathcal{L} = [-1,1]^V$, $\mathcal{L}^{1/2} = \{-1,0,1\}^V$. It will be convenient to work with functions $\hat{h} : \mathcal{L} \to \mathbb{R}$ and $h : \mathcal{L}^{1/2} \to \mathbb{R}$ obtained from $\hat{f}$ and $f$ via a linear change of coordinates $x_i \mapsto 2x_i - 1$. Under this change totally half-integral relaxations are transformed to *totally integral* relaxations:

**Definition 4.** *Let $\hat{h} : \mathcal{L} \to \mathbb{R}$ be a function of $n$ variables. (a) $\hat{h}$ is called* integral *if it is a convex polyhedral function such that all extreme points of the epigraph $\{(\boldsymbol{x}, z) \mid \boldsymbol{x} \in \mathcal{L}, z \geq \hat{h}(\boldsymbol{x})\}$ have the form $(\boldsymbol{x}, \hat{h}(\boldsymbol{x}))$ where $\boldsymbol{x} \in \mathcal{L}^{1/2}$. (b) $\hat{h}$ is called* totally integral *if it is integral and for an arbitrary ordering of nodes the following functions of $n-1$ variables (if $n > 1$) are totally integral:*

$$\begin{aligned}
\hat{h}'(x_1, \ldots, x_{n-1}) &= \hat{h}(x_1, \ldots, x_{n-1}, x_{n-1}) \\
\hat{h}'(x_1, \ldots, x_{n-1}) &= \hat{h}(x_1, \ldots, x_{n-1}, -x_{n-1}) \\
\hat{h}'(x_1, \ldots, x_{n-1}) &= \hat{h}(x_1, \ldots, x_{n-1}, \gamma) && \text{for any constant } \gamma \in \{-1, 0, 1\}
\end{aligned}$$

The definition of a bisubmodular function is adapted as follows: function $h : \mathcal{L}^{1/2} \to \mathbb{R}$ is bisubmodular if inequality (1) holds for all $\boldsymbol{x}, \boldsymbol{y} \in \mathcal{L}^{1/2}$ where operations $\sqcap, \sqcup$ are defined by tables (2) after replacements $0 \mapsto -1$, $\frac{1}{2} \mapsto 0$, $1 \mapsto 1$. To prove theorem 3, it suffices to establish a link between totally integral relaxations $\hat{h} : \mathcal{L} \to \mathbb{R}$ and bisubmodular functions $h : \mathcal{L}^{1/2} \to \mathbb{R}$. We can assume without loss of generality that $\hat{h}(\mathbf{0}) = h(\mathbf{0}) = 0$, since adding a constant to the functions does not affect the theorem.

A pair $\omega = (\pi, \boldsymbol{\sigma})$ where $\pi : V \to \{1, \ldots, n\}$ is a permutation of $V$ and $\boldsymbol{\sigma} \in \{-1, 1\}^V$ will be called a *signed ordering*. Let us rename nodes in $V$ so that $\pi(i) = i$. To each signed ordering $\omega$ we associate labelings $\boldsymbol{x}^0, \boldsymbol{x}^1, \ldots, \boldsymbol{x}^n \in \mathcal{L}^{1/2}$ as follows:

$$\boldsymbol{x}^0 = (0, 0, \ldots, 0) \qquad \boldsymbol{x}^1 = (\sigma_1, 0, \ldots, 0) \qquad \ldots \qquad \boldsymbol{x}^n = (\sigma_1, \sigma_2, \ldots, \sigma_n) \tag{3}$$

where nodes are ordered according to $\pi$.

Consider function $h : \mathcal{L}^{1/2} \to \mathbb{R}$ with $h(\mathbf{0}) = 0$. Its *Lovász extension* $\hat{h} : \mathbb{R}^V \to \mathbb{R}$ is defined in the following way [28]. Given a vector $\boldsymbol{x} \in \mathbb{R}^V$, select a signed ordering $\omega = (\pi, \boldsymbol{\sigma})$ as follows: (i) choose $\pi$ so that values $|x_i|$, $i \in V$ are non-increasing, and rename nodes accordingly so that $|x_1| \geq \ldots \geq |x_n|$; (ii) if $x_i \neq 0$ set $\sigma_i = \mathtt{sign}(x_i)$, otherwise choose $\sigma_i \in \{-1, 1\}$ arbitrarily. It is not difficult to check that

$$\boldsymbol{x} = \sum_{i=1}^{n} \lambda_i \boldsymbol{x}^i \tag{4a}$$

where labelings $\boldsymbol{x}^i$ are defined in (3) (with respect to the selected signed ordering) and $\lambda_i = |x_i| - |x_{i+1}|$ for $i = 1, \ldots, n-1$, $\lambda_n = |x_n|$. The value of the Lovász extension is now defined as

$$\hat{h}(\boldsymbol{x}) = \sum_{i=1}^{n} \lambda_i h(\boldsymbol{x}^i) \tag{4b}$$

**Theorem 5** ([28])**.** *Function $h$ is bisubmodular if and only if its Lovász extension $\hat{h}$ is convex on $\mathcal{L}$.* [2]

Let $\mathcal{L}_\omega$ be the set of vectors in $\mathcal{L}$ for which signed ordering $\omega = (\pi, \boldsymbol{\sigma})$ can be selected. Clearly, $\mathcal{L}_\omega = \{\boldsymbol{x} \in \mathcal{L} \mid |x_1| \geq \ldots \geq |x_n|,\ x_i\sigma_i \geq 0\ \forall i \in V\}$. It is easy to check that $\mathcal{L}_\omega$ is the convex hull of $n+1$ points (3). Equations (4) imply that $\hat{h}$ is linear on $\mathcal{L}_\omega$ and coincides with $h$ in each corner $x^0, \ldots, x^n$.

**Lemma 6.** *Suppose function $\tilde{h} : \mathcal{L} \to \mathbb{R}$ is totally integral. Then $\tilde{h}$ is linear on simplex $\mathcal{L}_\omega$ for each signed ordering $\omega = (\pi, \boldsymbol{\sigma})$.*

*Proof.* We use induction on $n = |V|$. For $n = 1$ the claim is straightforward; suppose that $n \geq 2$. Consider signed ordering $\omega = (\pi, \boldsymbol{\sigma})$. We need to prove that $\tilde{h}$ is linear on the boundary $\partial\mathcal{L}_\omega$; this will imply that $\hat{g}$ is linear on $\mathcal{L}_\omega$ since otherwise $\tilde{h}$ would have an extreme point in the the interior $\mathcal{L}_\omega \backslash \partial\mathcal{L}_\omega$ which cannot be integral.

Let $X = \{\boldsymbol{x}^0, \ldots, \boldsymbol{x}^n\}$ be the set of extreme points of $\mathcal{L}_\omega$ defined by (3). The boundary $\partial\mathcal{L}_\omega$ is the union of $n+1$ facets $\mathcal{L}^0_\omega, \ldots, \mathcal{L}^n_\omega$ where $\mathcal{L}^i_\omega$ is the convex hull of points in $X \backslash \{\boldsymbol{x}^i\}$. Let us prove that $\tilde{h}$ is linear on $\mathcal{L}^0_\omega$. All points $\boldsymbol{x} \in X \backslash \{\boldsymbol{x}^0\}$ satisfy $x_1 = \sigma_1$, therefore $\mathcal{L}^0_\omega = \{\boldsymbol{x} \in \mathcal{L}_\omega \mid x_1 = \sigma_1\}$. Consider function of $n-1$ variables $\tilde{h}'(x_2, \ldots, x_n) = \tilde{h}(\sigma_1, x_2, \ldots, x_n)$, and let $\mathcal{L}'^0_\omega$ be the projection of $\mathcal{L}^0_\omega$ to $\mathbb{R}^{V \backslash \{1\}}$. By the induction hypothesis $\tilde{h}'$ is linear on $\mathcal{L}'^0_\omega$, and thus $\tilde{h}$ is linear on $\mathcal{L}^0_\omega$.

The fact that $\tilde{h}$ is linear on other facets can be proved in a similar way. Note that for $i = 2, \ldots, n-1$ there holds $\mathcal{L}^i_\omega = \{\boldsymbol{x} \in \mathcal{L}_\omega \mid x_i = \sigma_{i-1}\sigma_i x_{i-1}\}$, and for $i = n$ we have $\mathcal{L}^n_\omega = \{\boldsymbol{x} \in \mathcal{L}_\omega \mid x_n = 0\}$. $\square$

[2] Note, Qi formulates this result slightly differently: $\hat{h}$ is assumed to be convex on $\mathbb{R}^V$ rather than on $\mathcal{L}$. However, it is easy to see that convexity of $\hat{h}$ on $\mathcal{L}$ implies convexity of $\hat{h}$ on $\mathbb{R}^V$. Indeed, it can be checked that $\hat{h}$ is positively homogeneous, i.e. $\hat{h}(\gamma\boldsymbol{x}) = \gamma\hat{h}(\boldsymbol{x})$ for any $\gamma \geq 0$, $\boldsymbol{x} \in \mathbb{R}^V$. Therefore, for any $\boldsymbol{x}, \boldsymbol{y} \in \mathbb{R}^V$ and $\alpha, \beta \geq 0$ with $\alpha + \beta = 1$ there holds

$$\hat{h}(\alpha\boldsymbol{x} + \beta\boldsymbol{y}) = \frac{1}{\gamma}\hat{h}(\alpha\gamma\boldsymbol{x} + \beta\gamma\boldsymbol{y}) \leq \frac{\alpha}{\gamma}\hat{h}(\gamma\boldsymbol{x}) + \frac{\beta}{\gamma}\hat{h}(\gamma\boldsymbol{y}) = \alpha\hat{h}(\boldsymbol{x}) + \beta\hat{h}(\boldsymbol{y})$$

where the inequality in the middle follows from convexity of $\hat{h}$ on $\mathcal{L}$, assuming that $\gamma$ is a sufficiently small constant.

**Corollary 7.** *Suppose function $\tilde{h} : \mathcal{L} \to \mathbb{R}$ with $\tilde{h}(\mathbf{0}) = 0$ is totally integral. Let $h$ be the restriction of $\tilde{h}$ to $\mathcal{L}^{1/2}$ and $\hat{h}$ be the Lovász extension of $h$. Then $\tilde{h}$ and $\hat{h}$ coincide on $\mathcal{L}$.*

Theorem 5 and corollary 7 imply the first part of theorem 3. The second part will follow from

**Lemma 8.** *If $h : \mathcal{L}^{1/2} \to \mathbb{R}$ with $h(\mathbf{0}) = 0$ is bisubmodular then its Lovász extension $\hat{h} : \mathcal{L} \to \mathbb{R}$ is totally integral.*

*Proof.* We use induction on $n = |V|$. For $n = 1$ the claim is straightforward; suppose that $n \geq 2$. By theorem 5, $\hat{h}$ is convex on $\mathcal{L}$. Function $\hat{h}$ is integral since it is linear on each simplex $\mathcal{L}_\omega$ and vertices of $\mathcal{L}_\omega$ belong to $\mathcal{L}^{1/2}$. It remains to show that functions $\hat{h}'$ considered in definition 4 are totally integral. Consider the following functions $h' : \{-1, 0, 1\}^{V \setminus \{n\}} \to \mathbb{R}$:

$$
\begin{aligned}
h'(x_1, \ldots, x_{n-1}) &= h(x_1, \ldots, x_{n-1}, x_{n-1}) \\
h'(x_1, \ldots, x_{n-1}) &= h(x_1, \ldots, x_{n-1}, -x_{n-1}) \\
h'(x_1, \ldots, x_{n-1}) &= h(x_1, \ldots, x_{n-1}, \gamma)\ , \quad \gamma \in \{-1, 0, 1\}
\end{aligned}
$$

It can be checked that these functions are bisubmodular, and their Lovász extensions coincide with respective functions $\hat{h}'$ used in definition 4. The claim now follows from the induction hypothesis. □

# 3 A new characterization of bisubmodularity

In this section we give an alternative definition of bisubmodularity; it will be helpful later for describing a relationship to the roof duality. As is often done for bisubmodular functions, we will encode each half-integral value $x_i \in \{0, 1, \frac{1}{2}\}$ via two binary variables $(u_i, u_{i'})$ according to the following rules:

$$0 \leftrightarrow (0,1) \qquad\qquad 1 \leftrightarrow (1,0) \qquad\qquad \tfrac{1}{2} \leftrightarrow (0,0)$$

Thus, labelings in $\mathcal{K}^{1/2}$ will be represented via labelings in the set

$$\mathcal{X}^- = \{\boldsymbol{u} \in \{0,1\}^{\overline{V}} \mid (u_i, u_{i'}) \neq (1,1) \quad \forall\, i \in V\}$$

where $\overline{V} = \{i, i' \mid i \in V\}$ is a set with $2n$ nodes. The node $i'$ for $i \in V$ is called the "mate" of $i$; intuitively, variable $u_{i'}$ corresponds to the complement of $u_i$. We define $(i')' = i$ for $i \in V$. Labelings in $\mathcal{X}^-$ will be denoted either by a single letter, e.g. $\boldsymbol{u}$ or $\boldsymbol{v}$, or by a pair of letters, e.g. $(\boldsymbol{x}, \boldsymbol{y})$. In the latter case we assume that the two components correspond to labelings of $V$ and $\overline{V} \backslash V$, respectively, and the order of variables in both components match. Using this convention, the one-to-one mapping $\mathcal{X}^- \to \mathcal{K}^{1/2}$ can be written as $(\boldsymbol{x}, \boldsymbol{y}) \mapsto \frac{1}{2}(\boldsymbol{x} + \overline{\boldsymbol{y}})$. Accordingly, instead of function $f : \mathcal{K}^{1/2} \to \mathbb{R}$ we will work with the function $g : \mathcal{X}^- \to \mathbb{R}$ defined by

$$g(\boldsymbol{x}, \boldsymbol{y}) = f\left(\frac{\boldsymbol{x} + \overline{\boldsymbol{y}}}{2}\right) \tag{5}$$

Note that the set of integer labelings $\mathcal{B} \subset \mathcal{K}^{1/2}$ corresponds to the set $\mathcal{X}^\circ = \{\boldsymbol{u} \in \mathcal{X}^- \mid (u_i, u_{i'}) \neq (0,0)\}$, so function $g : \mathcal{X}^- \to \mathbb{R}$ can be viewed as a discrete relaxation of function $g : \mathcal{X}^\circ \to \mathbb{R}$.

**Definition 9.** *Function $f : \mathcal{X}^- \to \mathbb{R}$ is called bisubmodular if*

$$f(\boldsymbol{u} \sqcap \boldsymbol{v}) + f(\boldsymbol{u} \sqcup \boldsymbol{v}) \;\; \leq \;\; f(\boldsymbol{u}) + f(\boldsymbol{v}) \qquad\qquad \forall\, \boldsymbol{u}, \boldsymbol{v} \in \mathcal{X}^- \tag{6}$$

*where $\boldsymbol{u} \sqcap \boldsymbol{v} = \boldsymbol{u} \wedge \boldsymbol{v}$, $\boldsymbol{u} \sqcup \boldsymbol{v} = \mathrm{REDUCE}(\boldsymbol{u} \vee \boldsymbol{v})$ and $\mathrm{REDUCE}(\boldsymbol{w})$ is the labeling obtaining from $\boldsymbol{w}$ by changing labels $(w_i, w_{i'})$ from $(1,1)$ to $(0,0)$ for all $i \in V$.*

To describe a new characterization, we need to introduce some additional notation. We denote $\mathcal{X} = \{0,1\}^{\overline{V}}$ to be the set of all binary labelings of $\overline{V}$. For a labeling $\boldsymbol{u} \in \mathcal{X}$, define labeling $\boldsymbol{u}'$ by $(\boldsymbol{u}')_i = \overline{u}_{i'}$. Labels $(u_i, u_{i'})$ are transformed according to the rules

$$(0,1) \to (0,1) \qquad (1,0) \to (1,0) \qquad\qquad (0,0) \to (1,1) \qquad (1,1) \to (0,0) \tag{7}$$

Equivalently, this mapping can be written as $(\boldsymbol{x}, \boldsymbol{y})' = (\overline{\boldsymbol{y}}, \overline{\boldsymbol{x}})$. Note that $\boldsymbol{u}'' = \boldsymbol{u}$, $(\boldsymbol{u} \wedge \boldsymbol{v})' = \boldsymbol{u}' \vee \boldsymbol{v}'$ and $(\boldsymbol{u} \vee \boldsymbol{v})' = \boldsymbol{u}' \wedge \boldsymbol{v}'$ for $\boldsymbol{u}, \boldsymbol{v} \in \mathcal{X}$. Next, we define sets

$$\begin{aligned}
\mathcal{X}^- &= \{\boldsymbol{u} \in \mathcal{X} \mid \boldsymbol{u} \le \boldsymbol{u}'\} = \{\boldsymbol{u} \in \mathcal{X} \mid (u_i, u_i') \neq (1,1) \quad \forall i \in \overline{V}\} \\
\mathcal{X}^+ &= \{\boldsymbol{u} \in \mathcal{X} \mid \boldsymbol{u} \ge \boldsymbol{u}'\} = \{\boldsymbol{u} \in \mathcal{X} \mid (u_i, u_i') \neq (0,0) \quad \forall i \in \overline{V}\} \\
\mathcal{X}^\circ &= \{\boldsymbol{u} \in \mathcal{X} \mid \boldsymbol{u} = \boldsymbol{u}'\} = \{\boldsymbol{u} \in \mathcal{X} \mid (u_i, u_i') \in \{(0,1),(1,0)\} \quad \forall i \in \overline{V}\} = \mathcal{X}^- \cap \mathcal{X}^+ \\
\mathcal{X}^\star &= \mathcal{X}^- \cup \mathcal{X}^+
\end{aligned}$$

Clearly, $\boldsymbol{u} \in \mathcal{X}^-$ if and only if $\boldsymbol{u}' \in \mathcal{X}^+$. Also, any function $g : \mathcal{X}^- \to \mathbb{R}$ can be uniquely extended to a function $g : \mathcal{X}^\star \to \mathbb{R}$ so that the following condition holds:

$$g(\boldsymbol{u}') = g(\boldsymbol{u}) \qquad \forall\, \boldsymbol{u} \in \mathcal{X}^\star \tag{8}$$

**Proposition 10.** *Let $g : \mathcal{X}^\star \to \mathbb{R}$ be a function satisfying* (8)*. The following conditions are equivalent:*

*(a)* *$g$ is bisubmodular, i.e. it satisfies* (6)*.*

*(b)* *$g$ satisfies the following inequalities:*

$$g(\boldsymbol{u} \wedge \boldsymbol{v}) + g(\boldsymbol{u} \vee \boldsymbol{v}) \;\; \le \;\; g(\boldsymbol{u}) + g(\boldsymbol{v}) \qquad \text{if} \;\; \boldsymbol{u}, \boldsymbol{v}, \boldsymbol{u} \wedge \boldsymbol{v}, \boldsymbol{u} \vee \boldsymbol{v} \in \mathcal{X}^\star \tag{9}$$

*(c)* *$g$ satisfies those inequalities in* (6) *for which $\boldsymbol{u} = \boldsymbol{w} \vee \boldsymbol{e}^i$, $\boldsymbol{v} = \boldsymbol{w} \vee \boldsymbol{e}^j$ where $\boldsymbol{w} = \boldsymbol{u} \wedge \boldsymbol{v}$ and $i, j$ are distinct nodes in $\overline{V}$ with $w_i = w_j = 0$. Here $\mathbf{e}^k$ for node $k \in \overline{V}$ denotes the labeling in $\mathcal{X}$ with $e^k_k = 1$ and $e^k_{k'} = 0$ for $k' \in \overline{V} \backslash \{k\}$.*

*(d)* *$g$ satisfies those inequalities in* (9) *for which $\boldsymbol{u} = \boldsymbol{w} \vee \boldsymbol{e}^i$, $\boldsymbol{v} = \boldsymbol{w} \vee \boldsymbol{e}^j$ where $\boldsymbol{w} = \boldsymbol{u} \wedge \boldsymbol{v}$ and $i, j$ are distinct nodes in $\overline{V}$ with $z_i = z_j = 0$.*

A proof is given in Appendix A. Note, an equivalent of characterization (c) was given by Ando et al. [2]; we state it here for completeness.

**Remark 1** We reformulated the bisubmodularity condition using standard operations $\wedge, \vee : \mathcal{X} \times \mathcal{X} \to \mathcal{X}$. This will be important in the next section for making a connection to the roof duality relaxation, which also uses operations $\wedge, \vee$. It is worth noting that set $\mathcal{X}^\star \subset \mathcal{X}$ is not closed under $\wedge, \vee$. (If $\mathcal{X}^\star$ were closed under $\wedge, \vee$ then (9) would be a definition of a submodular function on a distributive lattice.)

**Remark 2** In order to compare characterizations (b,d) to existing characterizations (a,c), we need to analyze the sets of inequalities in (b,d) **modulo eq.** (8), i.e. after replacing terms $g(\boldsymbol{w})$, $\boldsymbol{w} \in \mathcal{X}^+$ with $g(\boldsymbol{w}')$. In can be seen that the inequalities in (a) are neither subset nor superset of those in (b)[3], so (b) is a new characterization. It is also possible to show that from this point of view (c) and (d) are equivalent.

[3]Denote $\boldsymbol{u} = \begin{pmatrix} 1&0&1&0 \\ 0&0&0&0 \end{pmatrix}$ and $\boldsymbol{v} = \begin{pmatrix} 0&1&0&0 \\ 0&0&1&0 \end{pmatrix}$ where the top and bottom rows correspond to the labelings of $V$ and $\overline{V} \backslash V$ respectively, with $|V| = 4$. Plugging pair $(\boldsymbol{u}, \boldsymbol{v})$ into (6) gives the following inequality:

$$g\begin{pmatrix} 0&0&0&0 \\ 0&0&0&0 \end{pmatrix} + g\begin{pmatrix} 1&1&0&0 \\ 0&0&0&0 \end{pmatrix} \le g\begin{pmatrix} 1&0&1&0 \\ 0&0&0&0 \end{pmatrix} + g\begin{pmatrix} 0&1&0&0 \\ 0&0&1&0 \end{pmatrix}$$

This inequality is a part of (a), but it is not present in (b): pairs $(\boldsymbol{u}, \boldsymbol{v})$ and $(\boldsymbol{u}', \boldsymbol{v}')$ do not satisfy the RHS of (9), while pairs $(\boldsymbol{u}, \boldsymbol{v}')$ and $(\boldsymbol{u}', \boldsymbol{v})$ give a different inequality:

$$g\begin{pmatrix} 1&0&0&0 \\ 0&0&0&0 \end{pmatrix} + g\begin{pmatrix} 0&1&0&0 \\ 0&0&0&0 \end{pmatrix} \le g\begin{pmatrix} 1&0&1&0 \\ 0&0&0&0 \end{pmatrix} + g\begin{pmatrix} 0&1&0&0 \\ 0&0&1&0 \end{pmatrix}$$

where we used condition (8). Conversely, the second inequality is a part of (b) but it is not present in (a).

## 4 Submodular relaxations and roof duality

Consider a submodular function $g : \mathcal{X} \to \mathbb{R}$ satisfying the following "symmetry" condition:

$$g(\boldsymbol{u}') = g(\boldsymbol{u}) \qquad \forall\, \boldsymbol{u} \in \mathcal{X} \tag{10}$$

We call such function $g$ a *submodular relaxation* of function $f(\boldsymbol{x}) = g(\boldsymbol{x}, \overline{\boldsymbol{x}})$. Clearly, it satisfies conditions of proposition 10, so $g$ is also a bisubmodular relaxation of $f$. Furthermore, minimizing $g$ is equivalent to minimizing its restriction $g : \mathcal{X}^- \to \mathbb{R}$; indeed, if $\boldsymbol{u} \in \mathcal{X}$ is a minimizer of $g$ then so are $\boldsymbol{u}'$ and $\boldsymbol{u} \wedge \boldsymbol{u}' \in \mathcal{X}^-$.

In this section we will do the following: (i) prove that any pseudo-boolean function $f : \mathcal{B} \to \mathbb{R}$ has a submodular relaxation $g : \mathcal{X} \to \mathbb{R}$; (ii) show that the roof duality relaxation for quadratic pseudo-boolean functions is a submodular relaxation, and it dominates all other bisubmodular relaxations; (iii) show that for non-quadratic pseudo-boolean functions bisubmodular relaxations can be tighter than submodular ones; (iv) prove that similar to the roof duality relaxation, bisubmodular relaxations possess the *persistency* property.

**Review of roof duality** Consider a quadratic pseudo-boolean function $f : \mathcal{B} \to \mathbb{R}$:

$$f(\boldsymbol{x}) = \sum_{i \in V} f_i(x_i) + \sum_{(i,j) \in E} f_{ij}(x_i, x_j) \tag{11}$$

where $(V, E)$ is an undirected graph and $x_i \in \{0, 1\}$ for $i \in V$ are binary variables. Hammer, Hansen and Simeone [13] formulated several linear programming relaxations of this function and showed their equivalence. One of these formulations was called a *roof dual*. An efficient maxflow-based method for solving the roof duality relaxation was given by Hammer, Boros and Sun [5, 4].

We will rely on this algorithmic description of the roof duality approach [4]. The method's idea can be summarized as follows. Each variable $x_i$ is replaced with two binary variables $u_i$ and $u_{i'}$ corresponding to $x_i$ and $1 - x_i$ respectively. The new set of nodes is $\overline{V} = \{i, i' \mid i \in V\}$. Next, function $f$ is transformed to a function $g : \mathcal{X} \to \mathbb{R}$ by replacing each term according to the following rules:

$$f_i(x_i) \;\mapsto\; \frac{1}{2}[f_i(u_i) + f_i(\overline{u}_{i'})] \tag{12a}$$

$$f_{ij}(x_i, x_j) \;\mapsto\; \frac{1}{2}[f_{ij}(u_i, u_j) + f_{ij}(\overline{u}_{i'}, \overline{u}_{j'})] \qquad \text{if } f_{ij}(\cdot,\cdot) \text{ is submodular} \tag{12b}$$

$$f_{ij}(x_i, x_j) \;\mapsto\; \frac{1}{2}[f_{ij}(u_i, \overline{u}_{j'}) + f_{ij}(\overline{u}_{i'}, u_j)] \qquad \text{if } f_{ij}(\cdot,\cdot) \text{ is not submodular} \tag{12c}$$

$g$ is a submodular quadratic pseudo-boolean function, so it can be minimized via a maxflow algorithm. If $\boldsymbol{u} \in \mathcal{X}$ is a minimizer of $g$ then the roof duality relaxation has a minimizer $\hat{\boldsymbol{x}}$ with $\hat{x}_i = \frac{1}{2}(u_i + \overline{u}_{i'})$ [4].

It is easy to check that $g(\boldsymbol{u}) = g(\boldsymbol{u}')$ for all $\boldsymbol{u} \in \mathcal{X}$, therefore $g$ is a submodular relaxation. Also, $f$ and $g$ are equivalent when $u_{i'} = \overline{u}_i$ for all $i \in V$, i.e.

$$g(\boldsymbol{x}, \overline{\boldsymbol{x}}) = f(\boldsymbol{x}) \qquad \forall \boldsymbol{x} \in \mathcal{B} \tag{13}$$

**Invariance to variable flipping** Suppose that $g$ is a (bi-)submodular relaxation of function $f : \mathcal{B} \to \mathbb{R}$. Let $i$ be a fixed node in $V$, and consider function $f'(\boldsymbol{x})$ obtained from $f(\boldsymbol{x})$ by a change of coordinates $x_i \mapsto \overline{x}_i$ and function $g'(\boldsymbol{u})$ obtained from $g(\boldsymbol{u})$ by swapping variables $u_i$ and $u_{i'}$. It is easy to check that $g'$ is a (bi-)submodular relaxation of $f'$. Furthermore, if $f$ is a quadratic pseudo-boolean function and $g$ is its submodular relaxation constructed by the roof duality approach, then applying the roof duality approach to $f'$ yields function $g'$. We will sometimes use such "flipping" operation for reducing the number of considered cases.

**Conversion to roof duality** Let us now consider a non-quadratic pseudo-boolean function $f : \mathcal{B} \to \mathbb{R}$. Several papers [32, 1, 16] proposed the following scheme: (1) Convert $f$ to a quadratic pseudo-boolean function $\tilde{f}$ by introducing $k$ auxiliary binary variables so that $f(\boldsymbol{x}) = \min_{\boldsymbol{\alpha} \in \{0,1\}^k} \tilde{f}(\boldsymbol{x}, \boldsymbol{\alpha})$ for all labelings $\boldsymbol{x} \in \mathcal{B}$. (2) Construct submodular relaxation $\tilde{g}(\boldsymbol{x}, \boldsymbol{\alpha}, \boldsymbol{y}, \boldsymbol{\beta})$ of $\tilde{f}$ by applying the roof duality relaxation to $\tilde{f}$; then

$$\tilde{g}(\boldsymbol{x}, \boldsymbol{\alpha}, \boldsymbol{y}, \boldsymbol{\beta}) = \tilde{g}(\overline{\boldsymbol{y}}, \overline{\boldsymbol{\beta}}, \overline{\boldsymbol{x}}, \overline{\boldsymbol{\alpha}}) \quad , \quad \tilde{g}(\boldsymbol{x}, \boldsymbol{\alpha}, \overline{\boldsymbol{x}}, \overline{\boldsymbol{\alpha}}) = \tilde{f}(\boldsymbol{x}, \boldsymbol{\alpha}) \qquad \forall \boldsymbol{x}, \boldsymbol{y} \in \mathcal{B}, \ \boldsymbol{\alpha}, \boldsymbol{\beta} \in \{0,1\}^k$$

(3) Obtain function $g$ by minimizing out auxiliary variables: $g(\boldsymbol{x}, \boldsymbol{y}) = \min_{\boldsymbol{\alpha}, \boldsymbol{\beta} \in \{0,1\}^k} \tilde{g}(\boldsymbol{x}, \boldsymbol{\alpha}, \boldsymbol{y}, \boldsymbol{\beta})$.

One can check that $g(\boldsymbol{x}, \boldsymbol{y}) = g(\overline{\boldsymbol{y}}, \overline{\boldsymbol{x}})$, so $g$ is a submodular relaxation[4]. In general, however, it may not be a relaxation of function $f$, i.e. (13) may not hold; we are only guaranteed to have $g(\boldsymbol{x}, \overline{\boldsymbol{x}}) \le f(\boldsymbol{x})$ for all labelings $\boldsymbol{x} \in \mathcal{B}$.

**Existence of submodular relaxations** It is easy to check that if $f : \mathcal{B} \to \mathbb{R}$ is submodular then function $g(\boldsymbol{x}, \boldsymbol{y}) = \frac{1}{2}[f(\boldsymbol{x}) + f(\overline{\boldsymbol{y}})]$ is a submodular relaxation of $f$.[5] Thus, monomials of the form $c\Pi_{i \in A} x_i$ where $c \le 0$ and $A \subseteq V$ have submodular relaxations. Using the "flipping" operation $x_i \mapsto \overline{x}_i$, we conclude that submodular relaxations also exist for monomials of the form $c\Pi_{i \in A} x_i \Pi_{i \in B} \overline{x}_i$ where $c \le 0$ and $A, B$ are disjoint subsets of $U$. It is known that any pseudo-boolean function $f$ can be represented as a sum of such monomials (see e.g. [4]; we need to represent $-f$ as a *posiform* and take its negative). This implies that any pseudo-boolean function $f$ has a submodular relaxation.

Note that this argument is due to Lu and Williams [24] who converted function $f$ to a sum of monomials of the form $c\Pi_{i \in A} x_i$ and $c\overline{x}_k \Pi_{i \in A} x_i$, $c \le 0$, $k \notin A$. It is possible to show that the relaxation proposed in [24] is equivalent to the submodular relaxation constructed by the scheme above (we omit the derivation).

**Submodular vs. bisubmodular relaxations** An important question is whether bisubmodular relaxations are more "powerful" compared to submodular ones. The next theorem gives a class of functions for which the answer is negative; its proof is given in Appendix B.

**Theorem 11.** *Let $g$ be the submodular relaxation of a quadratic pseudo-boolean function $f$ defined by* (12)*, and assume that the set $E$ does not have parallel edges. Then $g$ dominates any other bisubmodular relaxation $\bar{g}$ of $f$, i.e. $g(\boldsymbol{u}) \ge \bar{g}(\boldsymbol{u})$ for all $\boldsymbol{u} \in \mathcal{X}^-$.*

For non-quadratic pseudo-boolean functions, however, the situation can be different. In Appendix C we give an example of a function $f$ of $n = 4$ variables which has a tight bisubmodular relaxation $g$ (i.e. $g$ has a minimizer in $\mathcal{X}^\circ$), but all submodular relaxations are not tight.

**Persistency** Finally, we show that bisubmodular functions possess the *autarky* property, which implies *persistency*.

**Proposition 12.** *Let $f : \mathcal{K}^{1/2} \to \mathbb{R}$ be a bisubmodular function and $\boldsymbol{x} \in \mathcal{K}^{1/2}$ be its minimizer.*

**[Autarky]** *Let $\boldsymbol{y}$ be a labeling in $\mathcal{B}$. Consider labeling $\boldsymbol{z} = (\boldsymbol{y} \sqcup \boldsymbol{x}) \sqcup \boldsymbol{x}$. Then $\boldsymbol{z} \in \mathcal{B}$ and $f(\boldsymbol{z}) \le f(\boldsymbol{y})$.*

**[Persistency]** *Function $f : \mathcal{B} \to \mathbb{R}$ has a minimizer $\boldsymbol{x}^* \in \mathcal{B}$ such that $x^*_i = x_i$ for nodes $i \in V$ with integral $x_i$.*

*Proof.* It can be checked that $z_i = y_i$ if $x_i = \frac{1}{2}$ and $z_i = x_i$ if $x_i \in \{0,1\}$. Thus, $\boldsymbol{z} \in \mathcal{B}$. For any $\boldsymbol{w} \in \mathcal{K}^{1/2}$ there holds $f(\boldsymbol{w} \sqcup \boldsymbol{x}) \le f(\boldsymbol{w}) + [f(\boldsymbol{x}) - f(\boldsymbol{w} \sqcap \boldsymbol{x})] \le f(\boldsymbol{w})$. This implies that $f((\boldsymbol{y} \sqcup \boldsymbol{x}) \sqcup \boldsymbol{x}) \le f(\boldsymbol{y})$. Applying the autarky property to a labeling $\boldsymbol{y} \in \arg\min\{f(\boldsymbol{x}) \mid \boldsymbol{x} \in \mathcal{B}\}$ yields persistency. □

[4] It is well-known that minimizing variables out preserves submodularity. Indeed, suppose that $h(\boldsymbol{x}) = \min_{\boldsymbol{\alpha}} \tilde{h}(\boldsymbol{x}, \boldsymbol{\alpha})$ where $\tilde{h}$ is a submodular function. Then $h$ is also submodular since

$$h(\boldsymbol{x}) + h(\boldsymbol{y}) = \tilde{h}(\boldsymbol{x}, \boldsymbol{\alpha}) + \tilde{h}(\boldsymbol{y}, \boldsymbol{\beta}) \ge \tilde{h}(\boldsymbol{x} \wedge \boldsymbol{y}, \boldsymbol{\alpha} \wedge \boldsymbol{\beta}) + \tilde{h}(\boldsymbol{x} \vee \boldsymbol{y}, \boldsymbol{\alpha} \vee \boldsymbol{\beta}) \ge h(\boldsymbol{x} \wedge \boldsymbol{y}) + h(\boldsymbol{x} \vee \boldsymbol{y})$$

[5] In fact, it dominates all other bisubmodular relaxations $\bar{g} : \mathcal{X}^- \to \mathbb{R}$ of $f$. Indeed, consider labeling $(\boldsymbol{x}, \boldsymbol{y}) \in \mathcal{X}^-$. It can be checked that $(\boldsymbol{x}, \boldsymbol{y}) = \boldsymbol{u} \sqcap \boldsymbol{v} = \boldsymbol{u} \sqcup \boldsymbol{v}$ where $\boldsymbol{u} = (\boldsymbol{x}, \overline{\boldsymbol{x}})$ and $\boldsymbol{v} = (\overline{\boldsymbol{y}}, \boldsymbol{y})$, therefore $\bar{g}(\boldsymbol{x}, \boldsymbol{y}) \le \frac{1}{2}[\bar{g}(\boldsymbol{u}) + \bar{g}(\boldsymbol{v})] = \frac{1}{2}[f(\boldsymbol{x}) + f(\overline{\boldsymbol{y}})] = g(\boldsymbol{x}, \boldsymbol{y})$.

## 5 Conclusions and future work

We showed that bisubmodular functions provide a natural generalization of the roof duality approach to higher-order terms. This can be viewed as a non-submodular analogue of the fact that submodular functions generalize the $s$-$t$ minimum cut problem with non-negative weights to higher-order terms.

As mentioned in the introduction, this work has been motivated by computer vision applications that use functions of the form $f(\boldsymbol{x}) = \sum_C f_C(\boldsymbol{x})$. An important open question is how to construct bisubmodular relaxations $\hat{f}_C$ for individual terms. For terms of low order, e.g. with $|C| = 3$, this potentially could be done by solving a small linear program.

Another important question is how to minimize such functions. Algorithms in [12, 25] are unlikely to be practical for most vision problems, which typically have tens of thousands of variables. However, in our case we need to minimize a bisubmodular function which has a special structure: it is represented as a sum of low-order bisubmodular terms. We recently showed [20] that a sum of low-order **submodular** terms can be optimized more efficiently using maxflow-like techniques. We conjecture that similar techniques can be developed for bisubmodular functions as well.

## Appendix A: Proof of proposition 10 (definitions of bisubmodularity)

Directions (a)⇒(c) and (b)⇒(d) are trivial. Below we prove directions (b)⇒(a), (d)⇒(b) and (c)⇒(d). We use the following notation: for a labeling $\boldsymbol{u} \in \mathcal{X}$ and distinct nodes $i, j \in \overline{V}$ we denote $[\boldsymbol{u}]_i = (u_i, u_{i'})$, $[\boldsymbol{u}]_{ij} = (u_i, u_{i'}, u_j, u_{j'})$.

**Direction (b)⇒(a)** For labelings $\boldsymbol{u}, \boldsymbol{v} \in \mathcal{X}^-$ define $\boldsymbol{\alpha} = \boldsymbol{u} \wedge \boldsymbol{v}'$, $\boldsymbol{\beta} = \boldsymbol{u}' \wedge \boldsymbol{v}$. Clearly, $\boldsymbol{\alpha}, \boldsymbol{\beta} \in \mathcal{X}^-$. Also, $\boldsymbol{\alpha} \wedge \boldsymbol{\beta} = \boldsymbol{u} \sqcap \boldsymbol{v}$ and $\boldsymbol{\alpha} \vee \boldsymbol{\beta} = \boldsymbol{u} \sqcup \boldsymbol{v}$. We can write

$$\begin{aligned} g(\boldsymbol{u} \sqcap \boldsymbol{v}) + g(\boldsymbol{u} \sqcup \boldsymbol{v}) &= g(\boldsymbol{\alpha} \wedge \boldsymbol{\beta}) + g(\boldsymbol{\alpha} \vee \boldsymbol{\beta}) \ \leq \ g(\boldsymbol{\alpha}) + g(\boldsymbol{\beta}) \ = \ g(\boldsymbol{\alpha}) + g(\boldsymbol{\beta}') \\ &= g(\boldsymbol{u} \wedge \boldsymbol{v}') + g(\boldsymbol{u} \vee \boldsymbol{v}') \ \leq \ g(\boldsymbol{u}) + g(\boldsymbol{v}') \ = \ g(\boldsymbol{u}) + g(\boldsymbol{v}) \end{aligned}$$

where we used conditions (8) and (9). (It can be checked that all labelings involved belong to $\mathcal{X}^\star$.)

**Direction (d)⇒(b)** We show that all inequalities in (9) hold using induction on the Hamming distance $||\boldsymbol{u} - \boldsymbol{v}||_1 = \sum_{i \in V} |u_i - v_i|$ between $\boldsymbol{u}$ and $\boldsymbol{v}$. The base case $||\boldsymbol{u} - \boldsymbol{v}||_1 \leq 2$ is straightforward: if labelings $\boldsymbol{u}, \boldsymbol{v}, \boldsymbol{u} \vee \boldsymbol{v}, \boldsymbol{u} \wedge \boldsymbol{v} \in \mathcal{X}^\star$ are all distinct then (9) follows directly from condition (d), otherwise (9) is an equality.

Suppose $||\boldsymbol{u} - \boldsymbol{v}||_1 \geq 3$ and $\boldsymbol{u}, \boldsymbol{v}, \boldsymbol{u} \vee \boldsymbol{v}, \boldsymbol{u} \wedge \boldsymbol{v} \in \mathcal{X}^\star$. We can assume by symmetry that $u_i = 1, v_i = 0$ for at least two nodes $i \in V$. Among such nodes, let us choose $i$ as follows: (a) if there is such $i$ with $u_{i'} = 1$ pick this node; (b) otherwise if there is such $i$ with $v_{i'} = 1$ pick this node; (c) otherwise pick an arbitrary such node. Let $\tilde{\boldsymbol{u}}$ be the labeling obtained from $\boldsymbol{u}$ by switching the label of $i$ from $1$ to $0$. It can be checked that $\boldsymbol{u} \vee (\tilde{\boldsymbol{u}} \vee \boldsymbol{v}) = \boldsymbol{u} \vee \boldsymbol{v}$ and $\boldsymbol{u} \wedge (\tilde{\boldsymbol{u}} \vee \boldsymbol{v}) = \tilde{\boldsymbol{u}}$. There holds $||\boldsymbol{u} - (\tilde{\boldsymbol{u}} \vee \boldsymbol{v})||_1 = |\{j \in V \mid u_j < v_j\}| + 1 < |\{j \in V \mid u_j < v_j\}| + |\{j \in V \mid u_j > v_j\}| = ||\boldsymbol{u} - \boldsymbol{v}||_1$ and $||\tilde{\boldsymbol{u}} - \boldsymbol{v}||_1 = ||\boldsymbol{u} - \boldsymbol{v}||_1 - 1$, so by the induction hypothesis

$$g(\boldsymbol{u} \vee \boldsymbol{v}) - g(\boldsymbol{u}) \leq g(\tilde{\boldsymbol{u}} \vee \boldsymbol{v}) - g(\tilde{\boldsymbol{u}}) \leq g(\boldsymbol{v}) - g(\tilde{\boldsymbol{u}} \wedge \boldsymbol{v}) = g(\boldsymbol{v}) - g(\boldsymbol{u} \wedge \boldsymbol{v})$$

provided that all labelings involved belong to $\mathcal{X}^\star$. This fact is proven below.

Let us show that $\tilde{\boldsymbol{u}} \in \mathcal{X}^\star$. If $u_{i'} = 1$ then $[\boldsymbol{u}]_i = (1, 1)$. $\tilde{\boldsymbol{u}}$ is obtained from $\boldsymbol{u}$ by switching $[\boldsymbol{u}]_i$ from $(1, 1)$ to $(0, 1)$, so $\boldsymbol{u} \in \mathcal{X}^\star$ implies $\tilde{\boldsymbol{u}} \in \mathcal{X}^\star$. Suppose that $u_{i'} = 0$; this means that $i$ was *not* selected in rule (a). If $\boldsymbol{u} \notin \mathcal{X}^-$ then there exists $j \in V$ with $[\boldsymbol{u}]_j = (1, 1)$. Since case (a) was not "triggered", we must have $[\boldsymbol{v}]_j = (1, 1)$. But then $[\boldsymbol{u} \wedge \boldsymbol{v}]_{ij} = (0, 0, 1, 1)$ so $\boldsymbol{u} \wedge \boldsymbol{v} \notin \mathcal{X}^\star$ - a contradiction. Thus, $\boldsymbol{u} \in \mathcal{X}^-$ and so $\tilde{\boldsymbol{u}} \in \mathcal{X}^-$.

Let us now show that $\tilde{\boldsymbol{w}} = \tilde{\boldsymbol{u}} \vee \boldsymbol{v} \in \mathcal{X}^\star$. Clearly, $\tilde{\boldsymbol{w}}$ is obtained from $\boldsymbol{w} = \boldsymbol{u} \vee \boldsymbol{v}$ by switching label $w_i$ from 1 to 0. If $\tilde{w}_{i'} = 1$ then $[\boldsymbol{w}]_i = (1, 1)$ switches to $[\tilde{\boldsymbol{w}}]_i = (0, 1)$, so $\boldsymbol{w} \in \mathcal{X}^\star$ implies $\tilde{\boldsymbol{w}} \in \mathcal{X}^\star$. Suppose that $\tilde{w}_{i'} = 0$, then $u_{i'} = v_{i'} = 0$; this means that rules (a) and (b) were not "triggered". Let us prove that $\boldsymbol{w} \in \mathcal{X}^-$; this will imply $\tilde{\boldsymbol{w}} \in \mathcal{X}^-$. Suppose not, then there exist $j \in V$ with $[\boldsymbol{w}]_j = (1,1)$. The case $[\boldsymbol{v}]_j = (1,1)$ is impossible since then we would have $[\boldsymbol{v}]_{ij} = (0,0,1,1)$ and $\boldsymbol{v} \notin \mathcal{X}^\star$ - a contradiction. Thus, we can assume by symmetry that $v_{j'} = 0$, so $u_{j'} = 1$. We must have $u_j = 1$ or $v_j = 1$, which means that either rule (a) or (b) would be triggered - a contradiction.

**Direction (c)⇒(d)** Let $\boldsymbol{u}, \boldsymbol{v}$ be labelings with the properties of condition (d). Thus, $\boldsymbol{u} = \boldsymbol{w} \vee \boldsymbol{e}^i$, $\boldsymbol{v} = \boldsymbol{w} \vee \boldsymbol{e}^j$ and $w_i = w_j = 0$.

Suppose that $j = i'$, so $[\boldsymbol{w}]_i = (0,0)$. We must have $[\boldsymbol{w}]_k \in \{(0,1),(1,0)\}$ for all $k \in V \backslash \{i,j\}$, otherwise we would have either $\boldsymbol{u} \wedge \boldsymbol{v} \notin \mathcal{X}^\star$ or $\boldsymbol{u} \vee \boldsymbol{v} \notin \mathcal{X}^\star$. Therefore, $\boldsymbol{u}, \boldsymbol{v} \in \mathcal{X}^\circ$ and $g(\boldsymbol{u} \sqcup \boldsymbol{v}) = g(\boldsymbol{w}) = g(\boldsymbol{w}') = g(\boldsymbol{u} \vee \boldsymbol{v})$, so (9) follows from (6).

We now assume that $j \neq i$ and $j \neq i'$. We have $[\boldsymbol{w}]_{ij} = (0,?,0,?)$. Cases $[\boldsymbol{w}]_{ij} = (0,0,0,1)$ and $[\boldsymbol{w}]_{ij} = (0,1,0,0)$ are impossible since then either $\boldsymbol{u}$ or $\boldsymbol{v}$ would not belong to $\mathcal{X}^\star$. If $[\boldsymbol{w}]_{ij} = (0,0,0,0)$ then $\boldsymbol{u}, \boldsymbol{v} \in \mathcal{X}^-$ and $\boldsymbol{u} \sqcup \boldsymbol{v} = \boldsymbol{u} \vee \boldsymbol{v}$, so (9) follows from (6). It remains to consider the case $[\boldsymbol{w}]_{ij} = [\boldsymbol{w}']_{ij} = (0,1,0,1)$. Labeling $\boldsymbol{u}'$ is obtained from $\boldsymbol{w}'$ by switching the label of node $i'$ from 1 to 0. Similarly, $\boldsymbol{v}'$ is obtained from $\boldsymbol{w}'$ by switching the label of node $j'$ from 1 to 0. We have $[\boldsymbol{u}']_{ij} = (0,0,0,1)$, $[\boldsymbol{v}']_{ij} = (0,1,0,0)$, so $\boldsymbol{u}', \boldsymbol{v}' \in \mathcal{X}^-$. Furthermore, $\boldsymbol{u}' \sqcup \boldsymbol{v}' = \boldsymbol{u}' \vee \boldsymbol{v}'$. Therefore,

$$g(\boldsymbol{u} \vee \boldsymbol{v}) + g(\boldsymbol{u} \wedge \boldsymbol{v}) = g(\boldsymbol{u}' \wedge \boldsymbol{v}') + g(\boldsymbol{u}' \vee \boldsymbol{v}') = g(\boldsymbol{u}' \sqcap \boldsymbol{v}') + g(\boldsymbol{u}' \sqcup \boldsymbol{v}') \leq g(\boldsymbol{u}') + g(\boldsymbol{v}') = g(\boldsymbol{u}) + g(\boldsymbol{v})$$

# Appendix B: Proof of theorem 11

In order to simplify the proof (i.e. reduce the number of considered cases) we will use the "flipping" operation described in section 4.

For $\boldsymbol{u} \in \mathcal{X}$ we define sets $V_{00}[\boldsymbol{u}] = \{i \in V \mid (u_i, u_{i'}) = (0,0)\}$, $\overline{V}_{00}[\boldsymbol{u}] = \{i \in \overline{V} \mid (u_i, u_{i'}) = (0,0)\}$. In this appendix we denote $\boldsymbol{u}^i$ to be the labeling obtained from labeling $\boldsymbol{u} \in \mathcal{X}$ by setting the label of node $i \in \overline{V}$ to 1, i.e. $\boldsymbol{u}^i = \boldsymbol{u} \vee \boldsymbol{e}^i$. Similarly, we denote $\boldsymbol{u}^{ij} = \boldsymbol{u} \vee \boldsymbol{e}^i \vee \boldsymbol{e}^j$.

**Lemma 13.** *Suppose that $\boldsymbol{u} \in \mathcal{X}$ and $i, j$ are distinct nodes in $\overline{V}_{00}[\boldsymbol{u}]$ satisfying the following conditions: (i) if $i, j \in V$ then the term $f_{ij}(\cdot,\cdot)$ (if it exists) is non-submodular; (ii) if $i, j \in \overline{V} \backslash V$ then the term $f_{i'j'}(\cdot,\cdot)$ (if it exists) is non-submodular; (iii) if $i \in V, j \in \overline{V} \backslash V$ then the term $f_{ij'}(\cdot,\cdot)$ (if it exists) is submodular; (iv) if $i \in \overline{V} \backslash V, j \in V$ then the term $f_{i'j}(\cdot,\cdot)$ (if it exists) is submodular. Then*

$$g(\boldsymbol{u}) + g(\boldsymbol{u}^{ij}) = g(\boldsymbol{u}^i) + g(\boldsymbol{u}^j) \tag{14}$$

*Proof.* It suffices to prove the lemma in the case when function $f$ in eq. (11) has a single term; the general case will then follow by linearity. If this term does not involve nodes $i/i'$ and $j/j'$ then the claim is trivial since then $g(\boldsymbol{u})$ does not depend on $u_i$ or $u_j$. Thus, we consider terms involving at least one of the nodes $i, i', j, j'$.

Suppose that $j = i'$; without loss of generality we can assume that $i \in V$. If $f(\boldsymbol{x}) = f_i(x_i)$ then the LHS and the RHS of (14) equal $f_i(0) + f_i(1)$. If $f(\boldsymbol{x}) = f_{ik}(x_i, x_k)$ and term $f_{ik}(\cdot,\cdot)$ is submodular then the LHS and the RHS of (14) equal $\frac{1}{2}[f_{ik}(0, u_k) + f_{ik}(1, \overline{u}_{k'}) + f_{ik}(1, u_k) + f_{ik}(0, \overline{u}_{k'})]$. The case when $f(\boldsymbol{x}) = f_{ik}(x_i, x_k)$ and term $f_{ik}(\cdot,\cdot)$ is non-submodular can be reduced to the previous one by flipping node $k$.

Now suppose that $j \neq i'$. By assumption, $(u_i, u_{i'}, u_j, u_{j'}) = (0,0,0,0)$. Using flipping, we can ensure that $i, j \in V$. (Note that flipping $i$ and/or $j$ preserves conditions (i)-(iv).) Suppose $f(\boldsymbol{x})$ involves exactly one of the nodes $i, j$, say node $i$. If $f(\boldsymbol{x}) = f_i(x_i)$ then the LHS and the RHS of (14) equal $\frac{1}{2}[f_i(0) + 3f_i(1)]$. If $f(\boldsymbol{x}) = f_{ik}(x_i, x_k)$ and term $f_{ik}(\cdot,\cdot)$ is submodular then the LHS and the RHS of (14) equal $\frac{1}{2}[f_{ik}(0, u_k) + f_{ik}(1, \overline{u}_{k'}) + f_{ik}(1, u_k) + f_{ik}(1, \overline{u}_{k'})]$. The case when $f(\boldsymbol{x}) = f_{ik}(x_i, x_k)$ and term $f_{ik}(\cdot,\cdot)$ is non-submodular can be reduced to the previous one by flipping node $k$. It remains to consider the case when $f(\boldsymbol{x}) = f_{ij}(x_i, x_j)$. By the lemma's assumption, term $f_{ij}(\cdot,\cdot)$ is non-submodular, so the LHS and the RHS of (14) equal $\frac{1}{2}[f_{ij}(0,1) + f_{ij}(1,0) + 2f_{ij}(1,1)]$. □

| | | | |
|---|---|---|---|
| $F(0,3)$ | | | |
| $F(0,2)$ | $F(1,2)$ | | |
| $F(0,1)$ | $F(1,1)$ | $F(2,1)$ | |
| $F(0,0)$ | $F(1,0)$ | $F(2,0)$ | $F(3,0)$ |

(a)

| | | | |
|---|---|---|---|
| $0$ | | | |
| $0$ | $0$ | | |
| $0$ | $0$ | $0$ | |
| $-1$ | $0$ | $1$ | $2$ |

(b)

| | | | |
|---|---|---|---|
| $0$ | | | |
| $-\frac{1}{3}$ | $0$ | | |
| $-\frac{2}{3}$ | $0$ | $0$ | |
| $-1$ | $0$ | $1$ | $2$ |

(c)

| | | | |
|---|---|---|---|
| 0000 | 3 | 1000 | 1 |
| 0001 | 2 | 1001 | 3 |
| 0010 | 4 | 1010 | 0 |
| 0011 | 10 | 1011 | 12 |
| 0100 | 2 | 1100 | 7 |
| 0101 | 12 | 1101 | 10 |
| 0110 | 13 | 1110 | 12 |
| 0111 | 12 | 1111 | 14 |

(d)

Figure 1: Examples of bisubmodular functions. (a-c) *Cardinality-dependent* functions $g : \mathcal{X}^- \to \mathbb{R}$ written as $g(\boldsymbol{u}) = G(n_{01}[\boldsymbol{u}], n_{10}[\boldsymbol{u}])$ where $n_{\alpha\beta}[\boldsymbol{u}] = |\{i \in V \mid (u_i, u_{i'}) = (\alpha, \beta)\}|$. Here $n = 3$. (a) Convention for displaying function $G$. (b,c) Bisubmodular relaxations of the same function $f$. Function (c) can be extended to a submodular relaxation, while (b) cannot be extended. (d) Function $f$ of $n = 4$ variables which has a tight bisubmodular relaxation, but all submodular relaxations are not tight.

**Lemma 14.** *Labeling $\boldsymbol{u} \in \mathcal{X}^-$ with $V_{00}[\boldsymbol{u}] = \{i\}$ satisfies $\bar{g}(\boldsymbol{u}) \le g(\boldsymbol{u})$.*

*Proof.* We have $2\bar{g}(\boldsymbol{u}) \overset{(1)}{\le} \bar{g}(\boldsymbol{u}^i) + \bar{g}(\boldsymbol{u}^{i'}) \overset{(2)}{=} g(\boldsymbol{u}^i) + g(\boldsymbol{u}^{i'}) \overset{(3)}{=} g(\boldsymbol{u}) + g(\boldsymbol{u}^{ii'}) \overset{(4)}{=} 2g(\boldsymbol{u})$ where (1) holds since $\bar{g}$ is bisubmodular and $\boldsymbol{u}^i \sqcap \boldsymbol{u}^{i'} = \boldsymbol{u}^i \sqcup \boldsymbol{u}^{i'} = \boldsymbol{u}$, (2) holds since $g$ and $\bar{g}$ are relaxations of $f$ and $\boldsymbol{u}^i, \boldsymbol{u}^{i'} \in \mathcal{X}^\circ$, (3) holds by lemma 13, and (4) holds since $g(\boldsymbol{u}^{ii'}) = g(\boldsymbol{u}') = g(\boldsymbol{u})$. □

**Lemma 15.** *There holds $\bar{g}(\boldsymbol{u}) - \bar{g}(\boldsymbol{u}^i) \le g(\boldsymbol{u}) - g(\boldsymbol{u}^i)$ for all $\boldsymbol{u} \in \mathcal{X}^-$ and $i \in \overline{V}_{00}[\boldsymbol{u}]$.*

*Proof.* We use induction on $|\overline{V}_{00}[\boldsymbol{u}]|$. If $\overline{V}_{00}[\boldsymbol{u}] = \varnothing$ then the claim is trivial. If $\overline{V}_{00}[\boldsymbol{u}] = \{i, i'\}$ then the claim follows from lemma 14 and the fact that $\bar{g}(\boldsymbol{u}^i) = g(\boldsymbol{u}^i)$ which holds since $\boldsymbol{u}^i \in \mathcal{X}^\circ$. Now suppose that there exists $j \in \overline{V}_{00}[\boldsymbol{u}] \backslash \{i, i'\}$. We can assume without loss of generality that labeling $\boldsymbol{u}$ and nodes $i, j$ satisfy conditions of lemma 13. (If not, we can replace $j$ with $j'$). We can write

$$\bar{g}(\boldsymbol{u}) - \bar{g}(\boldsymbol{u}^i) \overset{(1)}{\le} \bar{g}(\boldsymbol{u}^j) - \bar{g}(\boldsymbol{u}^{ij}) \overset{(2)}{\le} g(\boldsymbol{u}^j) - g(\boldsymbol{u}^{ij}) \overset{(3)}{=} g(\boldsymbol{u}) - g(\boldsymbol{u}^i)$$

where (1) holds since $\bar{g}$ is bisubmodular, (2) holds by the induction hypothesis and (3) follows from lemma 13. □

We are now ready to prove theorem 11, i.e. that $\bar{g}(\boldsymbol{u}) \le g(\boldsymbol{u})$ for any $\boldsymbol{u} \in \mathcal{X}^-$. We use induction on $|\overline{V}_{00}[\boldsymbol{u}]|$. The case $V_{00}[\boldsymbol{u}] = \varnothing$ is trivial and the case $V_{00}[\boldsymbol{u}] = \{i\}$ follows from lemma 14. Suppose that there exists $j \in \overline{V}_{00}[\boldsymbol{u}] \backslash \{i, i'\}$. As before, we can assume without loss of generality that labeling $\boldsymbol{u}$ and nodes $i, j$ satisfy conditions of lemma 13. We can write

$$\bar{g}(\boldsymbol{u}) \overset{(1)}{\le} \bar{g}(\boldsymbol{u}^i) + [\bar{g}(\boldsymbol{u}^i) - \bar{g}(\boldsymbol{u}^{ij})] \overset{(2)}{\le} g(\boldsymbol{u}^i) + [g(\boldsymbol{u}^i) - g(\boldsymbol{u}^{ij})] \overset{(3)}{=} g(\boldsymbol{u})$$

where (1) holds since $\bar{g}$ is bisubmodular, (2) holds by the induction hypothesis and lemma 14, and (3) follows from lemma 13.

## Appendix C: Examples of bisubmodular functions

First, let us consider *cardinality-dependent* functions $g : \mathcal{X}^- \to \mathbb{R}$, i.e. functions which can be expressed as

$$g(\boldsymbol{u}) = G(n_{01}[\boldsymbol{u}], n_{10}[\boldsymbol{u}])$$

where $n_{\alpha\beta}[\boldsymbol{u}] = |\{i \in V \mid (u_i, u_{i'}) = (\alpha, \beta)\}|$ for a labeling $\boldsymbol{u} \in \mathcal{X}$ and function $G$ is defined over $D_n = \{(a,b) \in \mathbb{Z}^2 \,|\, a, b \geq 0, a+b \leq n\}$. Using proposition 10(c) it is easy to check that $f$ is bisubmodular if and only if $G$ satisfies the following:

$$
\begin{aligned}
G(a,b) + G(a-2,b) &\leq 2G(a-1,b) && \forall (a,b) \in D_n, a \geq 2 && \text{(15a)} \\
G(a,b) + G(a,b-2) &\leq 2G(a,b-2) && \forall (a,b) \in D_n, b \geq 2 && \text{(15b)} \\
G(a,b) + G(a-1,b-1) &\leq G(a-1,b) + G(a,b-1) && \forall (a,b) \in D_n, a \geq 1, b \geq 1 && \text{(15c)} \\
2G(a,b) &\leq G(a+1,b+1) && \forall (a,b) \in D_n, a+b = n-1 && \text{(15d)}
\end{aligned}
$$

**Proposition 16.** *Consider function $G : D_3 \to \mathbb{R}$ defined by Figure 5(b). Let $g$ be the corresponding cardinality-dependent function of $2n = 6$ binary variables. Then (a) $g$ is bisubmodular and (b) it cannot be extended to a submodular relaxation $\mathcal{X} \to \mathbb{R}$.*

*Proof.* Verifying that function $G$ satisfies (15) is straightforward, so we focus on the claim (b). Suppose a submodular relaxation $\bar{g} : \mathcal{X} \to \mathbb{R}$ that extends $g$ does exist. Without loss of generality we can assume that $\bar{g}(\boldsymbol{u}) = \bar{G}(n_{01}[\boldsymbol{u}], n_{10}[\boldsymbol{u}], n_{00}[\boldsymbol{u}], n_{11}[\boldsymbol{u}])$ for some function $\bar{G}$ over $\bar{D}_3 = \{(a,b,c,d) \in \mathbb{Z}^2 \mid a,b,c,d \geq 0, a+b+c+d = 3\}$. Indeed, let $\Pi$ be the set of $3!=6$ permutations of $V$. Any permutation $\pi \in \Pi$ defines a mapping $\psi_\pi : \mathcal{X} \to \mathcal{X}$ in a natural way. Define function $\bar{g}_\pi : \mathcal{X} \to \mathbb{R}$ by $\bar{g}_\pi(\boldsymbol{u}) = \bar{g}(\psi_\pi(\boldsymbol{u}))$. Clearly, $\bar{g}_\pi$ is also a submodular relaxation extending $g$, and therefore so is the function $\bar{g}_* : \mathcal{X} \to \mathbb{R}$ given by $\bar{g}_*(\boldsymbol{u}) = \frac{1}{|\Pi|} \sum_{\pi \in \Pi} \bar{g}_\pi(\boldsymbol{u})$. Clearly, $\bar{g}_*(\boldsymbol{u})$ depends only on the counts $n_{01}[\boldsymbol{u}], n_{10}[\boldsymbol{u}], n_{00}[\boldsymbol{u}], n_{11}[\boldsymbol{u}]$.

If $c = 0$ or $d = 0$ for $(a,b,c,d) \in \bar{D}_3$ then $\bar{G}(a,b,c,d) = G(a,b)$. Thus, there are 4 unknown values: $\bar{G}(0,1,1,1)$, $\bar{G}(1,0,1,1)$, $\bar{G}(0,0,1,2)$, $\bar{G}(0,0,2,1)$. We will write labelings in $\mathcal{X}$ as $\left(\begin{smallmatrix} u_i & u_j & u_k \\ u_{i'} & u_{j'} & u_{k'} \end{smallmatrix}\right)$ where $\{i,j,k\} = U$. From submodularity of $\bar{g}$ we get

$$
\begin{aligned}
\bar{g}\left(\begin{smallmatrix} 1&0&1 \\ 0&0&1 \end{smallmatrix}\right) + \bar{g}\left(\begin{smallmatrix} 1&1&1 \\ 0&1&1 \end{smallmatrix}\right) \leq \bar{g}\left(\begin{smallmatrix} 1&0&1 \\ 0&1&1 \end{smallmatrix}\right) + \bar{g}\left(\begin{smallmatrix} 1&1&1 \\ 0&0&1 \end{smallmatrix}\right) \quad &\Rightarrow \quad \bar{G}(0,1,1,1) + 0 \leq 0 + 0 \\
\bar{g}\left(\begin{smallmatrix} 1&0&1 \\ 0&0&0 \end{smallmatrix}\right) + \bar{g}\left(\begin{smallmatrix} 1&0&1 \\ 0&1&1 \end{smallmatrix}\right) \leq \bar{g}\left(\begin{smallmatrix} 1&0&1 \\ 0&0&1 \end{smallmatrix}\right) + \bar{g}\left(\begin{smallmatrix} 1&0&1 \\ 0&1&0 \end{smallmatrix}\right) \quad &\Rightarrow \quad 0 + 0 \leq \bar{G}(0,1,1,1) + 0 \\
\bar{g}\left(\begin{smallmatrix} 0&0&1 \\ 1&0&1 \end{smallmatrix}\right) + \bar{g}\left(\begin{smallmatrix} 0&1&1 \\ 1&1&1 \end{smallmatrix}\right) \leq \bar{g}\left(\begin{smallmatrix} 0&0&1 \\ 1&1&1 \end{smallmatrix}\right) + \bar{g}\left(\begin{smallmatrix} 0&1&1 \\ 1&0&1 \end{smallmatrix}\right) \quad &\Rightarrow \quad \bar{G}(1,0,1,1) + 0 \leq 1 + 0 \\
\bar{g}\left(\begin{smallmatrix} 0&0&0 \\ 1&0&1 \end{smallmatrix}\right) + \bar{g}\left(\begin{smallmatrix} 0&1&1 \\ 1&0&1 \end{smallmatrix}\right) \leq \bar{g}\left(\begin{smallmatrix} 0&0&1 \\ 1&0&1 \end{smallmatrix}\right) + \bar{g}\left(\begin{smallmatrix} 0&1&0 \\ 1&0&1 \end{smallmatrix}\right) \quad &\Rightarrow \quad 1 + 0 \leq \bar{G}(1,0,1,1) + 0
\end{aligned}
$$

which implies $\bar{G}(0,1,1,1) = 0$, $\bar{G}(1,0,1,1) = 1$. Additional submodularity inequalities lead to an inconsistency:

$$
\begin{aligned}
\bar{g}\left(\begin{smallmatrix} 0&0&1 \\ 0&1&1 \end{smallmatrix}\right) + \bar{g}\left(\begin{smallmatrix} 1&1&1 \\ 0&1&1 \end{smallmatrix}\right) \leq \bar{g}\left(\begin{smallmatrix} 0&1&1 \\ 0&1&1 \end{smallmatrix}\right) + \bar{g}\left(\begin{smallmatrix} 1&0&1 \\ 0&1&1 \end{smallmatrix}\right) \quad &\Rightarrow \quad 1 + 0 \leq \bar{G}(0,0,1,2) + 0 \\
\bar{g}\left(\begin{smallmatrix} 0&1&1 \\ 0&0&0 \end{smallmatrix}\right) + \bar{g}\left(\begin{smallmatrix} 0&1&1 \\ 0&1&1 \end{smallmatrix}\right) \leq \bar{g}\left(\begin{smallmatrix} 0&1&1 \\ 0&0&1 \end{smallmatrix}\right) + \bar{g}\left(\begin{smallmatrix} 0&1&1 \\ 0&1&0 \end{smallmatrix}\right) \quad &\Rightarrow \quad 0 + \bar{G}(0,0,1,2) \leq 0 + 0
\end{aligned}
$$

□

It should be said that in this particular example $f$ has a submodular relaxation $\widetilde{g}$ with the same minimum as $g$ (the restriction of $\widetilde{g}$ to $\mathcal{X}^-$ is shown in Figure 5(c)). Although $\widetilde{g}(\boldsymbol{u}) < g(\boldsymbol{u})$ for some $\boldsymbol{u} \in \mathcal{X}^-$, both functions attain the minimum of $-1$ at $\boldsymbol{u} = \mathbf{0}$. Using a computer implementation with an exact rational LP solver `QSopt` [1] we found other examples of functions $f$ with $n = 4$ variables which have tight bisubmodular relaxations $g$ (i.e. $g$ has a minimizer in $\mathcal{X}^\circ$), but all submodular relaxations are not tight. One such example is shown in Figure 5(d); in this example, the minima of the tightest bisubmodular and submodular relaxations are $0$ and $-3/10$, respectively.